\documentclass[useAMS]{mn2e}
\usepackage{epsfig}
\usepackage{color}

\title[M31 globular cluster X-ray sources]{Demonstrating the likely neutron star
  nature of five M31 globular cluster sources with  Swift-NuSTAR spectroscopy}

\author[Maccarone et al.]{Thomas J. Maccarone$^1$\thanks{email: thomas.maccarone@ttu.edu}, Mihoko Yukita$^{2,3}$, Ann Hornschemeier$^3$, Bret D. Lehmer$^{2,3}$
\newauthor Vallia Antoniou$^4$, Andrew Ptak$^3$, Daniel R. Wik$^3$, Andreas Zezas$^5$,
\newauthor Padi Boyd$^3$, Jamie Kennea$^6$, Kim L. Page$^7$
\newauthor Mike Eracleous$^6$, Benjamin F. Williams$^8$, Steven E. Boggs$^9$
\newauthor Finn E. Christensen$^{10}$, William W. Craig$^9$, Charles J. Hailey$^{11}$, 
\newauthor Fiona A. Harrison$^{13}$, Daniel Stern$^{14}$, William W. Zhang$^3$
\\
$^1$Department of Physics, Box
  41051, Science Building, Texas Tech University, Lubbock TX, 
  79409-1051, USA\\
$^2$The Johns Hopkins University, Homewood Campus, Baltimore MD, 21218, USA\\
$^3$ NASA-Goddard Space Flight Center, Code 662, Greenbelt MD, 20771, USA\\
$^4$ Harvard-Smithsonian Center for Astrophysics, 60 Garden St, Cambridge, MA, 02138, USA\\
$^5$ Department of Physics \& Institute of Theoretical \& Computational Physics, University of Crete, 71003 Heraklion,  \\
Crete, Greece and Foundation for Research \& Technology-Hellas, 71110 Heraklion, Crete, Greece\\
$^6$ Department of Astronomy \& Astrophysics, The Pennsylvania State University,525 Davey Lab,  State College, PA, 16802, USA\\
$^7$ Department of Physics \& Astronomy, University of Leicester, Leicester LE2 2LL\\
$^8$ Department of Astronomy, University of Washington , Seattle, WA, USA\\
$^9$ University of California Space Sciences Laboratory, Berkeley, CA, USA\\
$^10$ National Space Institute, Technical University of Denmark, DK-2100, Copenhagen, Denmark\\
$^{11}$ Department of Physics, Columbia University, New York NY, USA\\
$^{12}$ Lawrence Livermore National Laboratory, Livermore CA, USA\\
$^{13}$ Division of Physics, Mathematics and Astronomy, California Institute of Technology, Pasadena CA, 91125, USA\\
$^{14}$ Jet Propulsion Laboratory, California Institute of Technology, Pasadena CA, 91109, USA\\
}

\begin{document}
\def\ltsim{\mathrel{\rlap{\lower 3pt\hbox{$\sim$}}
        \raise 2.0pt\hbox{$<$}}}
\def\gtsim{\mathrel{\rlap{\lower 3pt\hbox{$\sim$}}
        \raise 2.0pt\hbox{$>$}}}

\date{}

\pagerange{\pageref{firstpage}--\pageref{lastpage}} \pubyear{}

\maketitle

\label{firstpage}

\begin{abstract}
We present the results of a joint Swift-NuSTAR spectroscopy campaign
on M31.  We focus on the five brightest globular cluster X-ray sources
in our fields. Two of these had previously been argued to be black
hole candidates on the basis of apparent hard-state spectra at
luminosities above those for which neutron stars are in hard states.
We show that these two sources are likely to be Z-sources (i.e. low
magnetic field neutron stars accreting near their Eddington limits),
or perhaps bright atoll sources (low magnetic field neutron stars
which are just a bit fainter than this level) on the basis of
simultaneous Swift and NuSTAR spectra which cover a broader range of
energies.  These new observations reveal spectral curvature above 6-8
keV that would be hard to detect without the broader energy coverage
the NuSTAR data provide relative to Chandra and XMM-Newton.  We show
that the other three sources are also likely to be bright neutron star
X-ray binaries, rather than black hole X-ray binaries.  We discuss why
it should already have been realized that it was unlikely that these
objects were black holes on the basis of their being persistent
sources, and we re-examine past work which suggested that tidal
capture products would be persistently bright X-ray emitters.  We
discuss how this problem is likely due to neglecting disk winds in
older work that predict which systems will be persistent and which
will be transient.

\end{abstract}


\section{Introduction}
It has long been known that there are more X-ray binaries per unit
stellar mass in globular clusters than in field stellar populations
(Clark 1975).  The process by which X-ray binaries form in globular
clusters is different from X-ray binary formation processes in low
density field star populations.  In globular clusters, close binaries
are formed through interactions between stars, be they tidal captures
(Fabian et al. 1975), exchange encounters (Hills 1976), or direct
collisions (Verbunt \& Hut 1987).  As a result, the orbital period
distributions of the systems may be quite different from one another.

Whether black holes exist in globular clusters is a topic of great
importance for understanding the dynamical evolution of clusters
(e.g. Sippel \& Hurley 2013; Heggie \& Giersz 2014; Morscher et
al. 2015), and the formation of gravitational wave sources.  Black
holes in globular clusters are likely to have a different mass
distribution than those in field X-ray binaries, extending up to
higher masses{ , because the black holes in field X-ray binaries
  form predominantly through common envelope evolution (e.g. van den
  Heuvel 1983), while the black holes in cluster X-ray binaries may
  have formed in from single stars, or wide binary progenitors, and
  then entered binaries through tidal capture (Fabian et al. 1975) or
  exchange interactions (Hills 1976).}.  One can compare, for example,
the expected distribution of black hole masses from single star
evolution (Fryer \& Kalogera 2001) with the observed distribution from
X-ray binaries (\"Ozel et al. 2010; Farr et al. 2011). { The
  observed black holes are lighter than the distribution predicted in
  the case of single star evolution, lending credence to the idea that
  common envelopes lead to lower black hole masses}.  There have {
  also} been suggestions that heavier black holes may also form at low
metallicity (e.g. Linden et al. 2010; Mapelli et al. 2010), { and
  many globular clusters are significantly more metal poor than the
  typical star in the Galactic field}.  { The overall level of
  X-ray emission, and the luminosities of the brightest individual
  X-ray sources are highest in the most metal poor star forming
  galaxies (Basu-Zych et al. 2013; Brorby et al. 2014) indicating that
  the metallicity must affect either the masses of the compact
  objects, or the number of close binaries with compact objects.}
These claims had { also} appeared to be supported by the reports of
$\sim30 M_\odot$ black holes in two low metallicty galaxies, in the
binaries IC~10~X-1 and NGC~300~X-1 (Prestwich et al. 2007; Crowther et
al. 2010).  The mass estimates for both of these objects have recently
been called into question because the phasing of the X-ray eclipses
relative to the radial velocity curves indicate that the radial
velocity curves are not tracing the orbits of the donor stars (Laycock
et al. 2015a,b; Binder et al. 2015).

There are a few more reasons why identifying stellar mass black hole
X-ray binaries in globular clusters is of major astrophysical
importance.  These objects are unlikely to survive in the same
globular clusters that contain intermediate mass black holes (Leigh et
al. 2014); instead, dynamical friction should cause them to sink to
the center of the cluster, where the IMBH will split the binaries.
Additionally, stellar mass black holes in globular clusters represent
an extreme case that can be used to test theories of space-times with
more than 4 dimensions in which Hawking radiation might be far more
efficient than in a spacetime described by standard general relativity
(Emparan et al. 2002; Psaltis 2007); globular clusters give excellent
``clocks'' for proving that the black hole in question is quite old,
so stellar mass black holes in globular clusters give the strongest
available constraints on this problem (Gnedin et al. 2009).

It has been suggested that the black hole X-ray binaries that form via
tidal capture should be persistent X-ray sources, while those that
form via exchange interactions might be predominantly transient
sources (Kalogera et al. 2004).  { We define the boundary between
  persistent and transient sources here to be sources which are
  unaffected and affected, respectively, by the ionization instability
  in their accretion disks -- i.e. persistent sources accrete rapidly
  enough that their outer accretion disks are ionized at all times,
  while transient sources are sources which have low enough accretion
  rates that they spend most of their times in states where the outer
  disk is neutral, and hence they are subject to this instability
  (e.g. Cannizzo, Wheeler \& Ghosh 1985; Cannizzo, Chen \& Livio 1995;
  King et al. 1996).  In practice, this should be associated with
  variations of a factor of $\sim10^4$ or more in luminosity, but
  given the long outbursts of sources with long orbital periods (see
  e.g. Truss \& Done 2006), there may be objects which appear to be
  persistent over the lifetime of X-ray astronomy, but which are
  undergoing such outburst cycles.}  The basis for { the suggestion
  that tidal capture sources would be persistent} comes from King et
al. (1996), where it was shown that black hole X-ray binaries with
orbital periods of a few to ten hours would typically have mass
transfer rates that would make them persistent sources.  Barnard et
al. (2008) use this as part of the argument for why it is reasonable
to find many persistent objects at luminosities of $10^{38}$ erg/sec
in M31 globular clusters, and to associate them with black hole
accretors.  On the other hand persistent black holes are not seen in
substantial numbers in the Galactic field populations.  The known
black hole X-ray binaries in this period range are predominantly
transient sources -- only one strong candidate black hole X-ray binary
with a low mass donor star is persistent -- 4U~1957+11 (Gomez et
al. 2015) -- { and even that object is not a dynamically confirmed
  black hole}.

For quite some time, it was thought that globular clusters would not
contain stellar mass black holes in substantial numbers.  Spitzer
(1969) had shown that dynamical decoupling would result based on a
criterion involving a critical combination of the fraction of the
cluster's mass, and the ratio of the masses of the heavy objects to
the masses of the light objects.  This criterion would be satisfied
for black holes in most old star clusters.  This then leads to a
combination of effects that should eject a large numbers of the black
holes -- dynamical evaporation and ejection in three body encounters
being the two most important (Kulkarni et al. 1993; Sigurdsson \&
Hernquist 1993).  Additionally, the gravitational radiation rocket
effect (Redmount \& Rees 1975) could also eject a large fraction of
any black holes that merge.

Additional discussion, both in the 1970's and in the past 15 years, has
concerned the possibility of finding intermediate mass black holes in
globular clusters.  In recent years, searches have been partially
motivated by placing globular clusters on the $M_{BH}-\sigma$ relation
for galaxies (Gebhardt et al. 2000;Ferrarese \& Merritt 2000) and
finding that, if the nature of the systems is the same, they should
host intermediate mass black holes; and partly by numerical
calculations that suggest that either mergers of stellar mass black
holes (Miller \& Hamilton 2002) or mergers of massive stars (Portegies
Zwart \& McMillan 2002) could lead to the production of intermediate
mass black holes in globular clusters.  Concrete observational
evidence has continued to be lacking.  Dynamical studies have, in some
cases, shown evidence for increasing mass-to-light ratios in the
centers of globular clusters (e.g. Newell et al. 1975; Gerssen et
al. 2002; Noyola et al. 2008).  Dynamics theory has argued that mass
segregation should place an excess of stellar remnants in the centers
of globular clusters (Illingworth \& King 1976; Baumgardt et
al. 2003).  Proper motion studies of Omega Cen to date have not shown
a need for an intermediate mass black hole (van der Marel \& Anderson
2010; Watkins et al. 2013).  Searches for accretion signatures, both
in X-rays (Grindlay et al. 2001; Haggard et al. 2013) and in radio
(e.g. Maccarone 2004; Strader et al. 2012a) have yielded only upper
limits, which in some cases are below the estimates from dynamical
studies.

On the other hand, over the past decade, the evidence for globular
clusters with stellar mass black holes has
mounted.  The first evidence was seen from extremely bright, strongly
variable sources in galaxies within 20 Mpc (Maccarone et al. 2007;
Brassington et al. 2008), followed by observations of ``ultrasoft''
spectra (White \& Marshall 1984) from moderately variable sources in
NGC~4472 (Maccarone et al. 2011).  More recently, flat spectrum radio
sources have been detected in the cores of many Milky Way clusters at
luminosities in excess of what is expected from neutron star X-ray
binaries (Strader et al. 2012b; Chomiuk et al. 2013).

In this paper, we use joint Swift-NuSTAR spectra of several bright
X-ray sources in M31 globular clusters to help determine whether they
are accreting black holes or accreting neutron stars.  The sources are
selected on the basis of being bright globular cluster X-ray sources
which are in our NuSTAR fields and are sufficiently isolated as to
allow straightforward spectroscopy.  Two of these have already been
claimed to be globular cluster black holes (Barnard et al. 2011) on
the basis of fits to spectra taken by Chandra and XMM-Newton.  In this
paper, we find that the spectra of both of those sources, as well as
those of two other bright globular cluster X-ray sources in M31, are
much better fit by models typically used to fit the spectra of neutron
stars than models typically used to fit the spectra of black holes.
We also discuss in this paper possible reasons why the prediction made
in Kalogera et al. (2004) that tidal capture products should be
persistent sources is at odds with observations of other Galactic
black hole X-ray binaries in a similar orbital period range.

\section{Spectral state phenomenology}

Accreting compact objects typically show a few key spectral states in
which substantial amounts of time are spent.  Historically the
nomenclature for these sources has been different for black holes and
neutron stars, but in recent years, terminology has begun to converge
for the lower luminosity, more stable source states.  

The first indications of spectral state dichotomy were discovered by
Tananbaum et al. (1972), who found, in Cygnus X-1, that the radio
emission turned off as the X-ray spectrum went from being dominated by
hard X-rays to being dominated by soft X-rays.  Hard states are well
modelled by thermal Comptonization in an optically thin, geometrically
thick hot flow (Thorne \& Price 1975).  These states are always seen
at low luminosities (in the ``low/hard states'', typically seen below
2\% of the Eddington limit -- Maccarone 2003), and are often seen at
higher luminosities at the starts of transient outbursts, due to a
hysteresis effect seen in black holes (Miyamoto et al. 1995) and found
to show analogous behavior in neutron stars (Maccarone \& Coppi 2003)
and even in accreting white dwarfs (Wheatley et al. 2003).  In the
accreting neutron stars, it was once common to refer to such states as
island states, following Hasinger \& van der Klis (1989), but in
recent years, the term ``hard state'' has been applied to both black
hole and neutron star accretion flows.

X-ray binaries also often exhibit states well explained by standard
accretion disk models { (e.g. Shakura \& Sunayev 1973; Davis et
  al. 2005)}, in which the emission is thermal with gravitational
energy release balanced by radiation.  These states are dominated by
soft X-rays, and are often called soft states.  Neutron stars with
similar accretion rates will typically show more complicated spectra,
presumably because there is emission from both the accretion disk and
the boundary layer (i.e. the region near the surface of the star where
the excess rotational energy of the inflow is dissipated) { -- see
  e.g. White \& Marshall (1984)}.  With high signal-to-noise ratio, it
is often necessary to use two components to model ``soft state''
neutron star spectra.  The spectra of neutron stars in such states
tend to peak at higher temperatures than the spectra of black holes,
but they still show strong curvature above 10 keV, rather than power
law spectra, and the difference in temperature is likely to be due
primarily to the $M^{-1/4}$ temperature dependence for accretion disks
at a constant Eddington fraction which extend in to the innermost
stable circular orbit.

Bright neutron stars often behave as ``Z-sources'', so named because
as they vary, they evolve through a colour-colour diagram along a path
that is shaped roughly like the letter ``Z'' (Hasinger \& van der Klis
1989). { The spectral shapes for these sources are not much
  different from those of the soft state sources.  They can generally
  be well-modelled by low temperature, moderate optical depth thermal
  Comptonization models when the count rates are low, and often
  require two quasi-thermal components when observed at high
  signal-to-noise ratio.  The brightest atoll sources -- i.e. the
  brightest ``soft state'' neutron stars -- have spectra that are
  quite difficult to distinguish from the Z-source spectra (e.g. Di
  Salvo et al. 2002; Gierli\'nski \& Done 2002), and there is even one
  source, XTE~J1701-462, which transitions between the atoll and Z
  behaviours, but which does not show any dramatic difference between
  Z-source aand bright atoll source spectra (Lin, Remillard \& Homan
  2009) .  The Z-sources are generally a bit more strongly variable
  than the brightest atoll sources (e.g. van der Klis 1995), but this
  distinction is not something of which we can take advantage when
  working with sources in M31 due to the relatively low count rates.}

Extremely bright black hole accretion disks, as well as black hole
accretion disks observed during the transition between the hard state
and the soft state, show different modes of behavior { (Miyamoto et
  al. 1991; Homan et al. 2001)}.  The brightest accretors probably
have radiation pressure-dominated disks { (Shakura \& Sunyaev
  1973)}, and can be moderately well modelled as steep power laws and
are sometimes called steep power law states (McClintock \& Remillard
2006).  These states are relatively uncommon and short-lived in most
systems.  They are seen fairly often in GRS~1915+105, but that object
is typically near the Eddington limit, and on the basis of its
luminosity alone, it would be classified as a black hole without much
debate, now that its distance is well established (Reid et
al. 2014).\footnote{ There is now an ultraluminous X-ray source,
  M82 X-2, which has been established to have a neutron star primary
  on the basis of pulsations (Bachetti et al. 2014).  This source
  shows a spectrum harder than that which is seem from black hole
  candidates at similar luminosities, and shows pulsations, both of
  which distinguish it from bright black hole X-ray binaries fairly
  clearly.  Its existence does suggest more caution on characterising
  sources solely based on luminosity, but it is quite
  phenomenologically different from black holes accreting above the
  neutron star Eddington limit, and the magnetic collimation that
  causes the pulsations to appear also probably allows the apparent
  luminosity from M82 X-2 to exceed the Eddington limit by such a
  large factor.}

\section{Observations}

We make use of data sets obtained from Swift and NuSTAR.  Both the
NuSTAR data and the Swift data have been extracted with 45'' apertures
around the already-known source positions (Galetti et al. 2004;
Peacock et al. 2010).  The names, positions and magnitudes of the
clusters are given in Table \ref{clusters}.  There do exist higher
signal-to-noise archival XMM and Chandra data for these sources, but
we prefer the quasi-simultaneous Swift data over the Chandra/XMM data,
because the systematic uncertainties that may be induced due to source
variability are hard to quantify and likely are more important than
the increased statistical errors from the Swift data.  The data are
grouped to a minimum of 1 count per bin to avoid some poorly
understood statistical problems.  In the plots, further rebinning is
done to help make the figures clearer, but these binnings are not used
for spectral fitting.  { Source-free background regions
  near the sources are used for creating background spectra for both
  instruments, and response matrices are generated with the standard
  tools for both satellites.  For NuSTAR, FMPA and FMPB data are
  combined, and an averaged response matrix is produced.}

We have three NuSTAR observations which were used for this project. These are listed in Table \ref{obstable}.

\begin{table*}
\begin{tabular}{lllllll}
Source& NuSTAR ObsID& NuSTAR dates& NuSTAR exposure& Swift ObsID& Swift dates & Swift exposure\\
\hline
Bo~153&50026001002&6-8 February 2015&106386&0008000700(1,2)&6,8 February 2015&17016\\
Bo~185&50026001002&6-8  February 2015&106386&0008000700(1,2)&6,8 February 2015&17016\\
Bo~225&50026002001&8-11 February 2015&108939&0008084600(1-3)& 8-11 February 2015&22996\\
Bo~375&50026003003&8-11 March 2015&104370&00080847003&8-9 March 2015&17311\\
SK182C&50026003003&8-11 March 2015&104370&00080847003&8-9 March 2015&17311\\
\end{tabular}
\caption{The observations used for this project.  The columns are: (1)
  the host globular cluster name (2) the observation ID number for the
  NuSTAR data (3) the dates for the NuSTAR observation (4) the total
  NuSTAR exposure time in seconds (5) the Swift observation ID
  number(s) (6) the date(s) for the Swift observations and (7) the
  total Swift exposure time in seconds.  There are two Swift
  observations for Bo~153 and Bo~185, and three Swift observations for
  Bo~225.}
\label{obstable}
\end{table*}

\section{Classification of compact object class based on X-ray data}

The gold standard for identifying black holes has traditionally been
demonstration that the mass of the accretor excedes the maximum mass
for a neutron star under equations of state allowed by both general
relativity and laboratory experiments on dense matter (Kalogera \&
Baym 1996).  The masses are typically estimated using a combination of
radial velocity curves, and some estimate of the binary inclination
angle (McClintock \& Remillard 1986; Casares \& Jonker 2014).  In many
cases, however, the distance, extinction and/or crowding make it
difficult or impossible to make a measurement of an object's radial
velocity curve.  Additionally, some sources are persistently bright,
making it impossible to estimate their inclination angles from
ellipsoidal modulations.

A variety of tests exists for showing that an accreting object is a neutron
star rather than a black hole.  The two most prominent are detection
of pulsations (Giacconi et al. 1971) and detection of Type I X-ray
bursts (first seen by Grindlay et al. 1976, but first associated with
thermonuclear fusion on a neutron star by Maraschi \& Cavaliere 1977
-- see also Woosley \& Taam 1976).  These phenomena can be used for
nearby sources in crowded or reddened regions, but typically do not
provide sufficiently strong signals to be detected in extragalactic
binaries, even in M31.  Additionally, the absence of bursts or
pulsations is rather difficult to use as strong evidence in favor of a
black hole.  There may be accretion regimes in which Type I bursts
would be expected if the object is a neutron star (e.g. Remillard et
al. 2006) such that strong indirect evidence would be provided.

At the same time, a phenomenology exists for demonstrating that an
object is a black hole rather than a neutron star, given more and
better X-ray data.  The origins of the ideas used date back to the
1980's, and have been fleshed out to the extent that they have reached
fairly wide acceptance, if not a total consensus.  White \& Marshall
(1984) suggested that the presence of an ultrasoft component in a
spectrum could be an indicator of a black hole rather than neutron
star accretor.  This suggestion has stood up well over time.
Gradually, it has been found that high/soft state black holes are well
modelled by a series of optically thick annuli with temperatures that
decrease outwards.  The disk blackbody model (\textsc{diskbb} in XSPEC
-- Mitsuda et al. 1984) provides an excellent phenomenological
description of the data.  There do exist more models which treat
the radiative transfer and relativistic effects in the disk in greater detail
(e.g. Davis et al. 2005) and have been used to estimate the inner disk
radii in order to make estimates of the spin of accreting black holes
(Zhang et al. 1997 for an early attempt; Shafee et al. 2006).  The
newer disk models provide much more precise parameter estimation, but
typically do not fit the data any better, and for the purposes of this
paper, in which we merely aim to classify the type of source spectrum,
the higher level of complication in using such models is not
justified.

The spectra of neutron stars are considerably more complex, and there
is less consensus about the correct models for describing the real
physics of the systems (see e.g. White et al. 1988; Mitsuda et
al. 1989; Church \& Baluci\'nska-Church 1995).  In this paper we will
use a simple thermal Comptonization model within XSPEC
(\textsc{comptt} -- Titarchuk 1994).  This model has been shown to
provide good spectral fits to bright accreting neutron stars in the
past (e.g. Lavagetto et al. 2008).\footnote{We are not particularly
  concerned with extracting detailed information about the spectra of
  the sources studied in this paper, given that Galactic and
  Magellanic Cloud sources will be better for that purpose.  We are
  primarily interested in understanding which sources are black holes
  and which are neutron stars.  We are thus concerned only about
  classification and hence choose a model with relatively few free
  parameters and which can parameterize the data well, rather than a
  model which is physically well-motivated.}  These sources typically
fit to relatively high optical depths ($\tau\sim10$) and low
temperature Comptonization ($k_BT \sim 3$ keV) models, with low
temperature seed photon distributions.

Additionally, state transitions from soft states to hard states occur
at a fairly uniform 2\% of the Eddington luminosity (Maccarone 2003;
Kalemci et al. 2013).\footnote{Dunn et al. (2010) suggested that there
  was as much spread in the soft-to-hard state transition luminosities
  as in the hard-to-soft state transition luminosities, but those
  claims were based entirely on including a set of objects without
  known black hole masses or distances, and assuming them to be at
  distances of 5 kpc (closer than the Galactic Center distance), and
  to have masses of 10 $M_\odot$ (larger than the typical 8 $M_\odot$
  value for other stellar mass black holes from \"Ozel et al. 2010;
  Farr et al. 2011)  The combination of these assumption
  systematically drives down the state transition luminosities for the
  poorly studied sources, creating a substantial amount of scatter
  which does not exist for the well-studied sources.}  Thus, if a
distance to a source is known, the state transition luminosity can be
used as an estimator of the compact object mass, which is sufficient
to distinguish between neutron stars of $\approx 1.4-2.0 M_\odot$ and
black holes of $5-10M_\odot$. Some hysteresis effects are seen in
black hole systems (Miyamoto et al. 1995) which are generally quite
similar to those seen in neutron star systems (Maccarone \& Coppi
2003), but the high luminosity hysteretic hard states are generally
quite short lived, and so are improbable to catch in a single
snapshot, and can be ruled out with monitoring observations.  Based on
an earlier understanding of black hole/neutron star phenomenology,
Barret et al. (1996) proposed that observing a source to have a hard
X-ray (i.e. $>$20 keV) luminosity above $10^{37}$ erg/sec was evidence
that a source is a black hole.

Single epoch X-ray spectroscopy can often separate out black holes
from neutron stars, as well.  In low/hard states, the spectra can
often be quite difficult to differentiate from one another, but
neutron stars often show cutoffs at somewhat lower energies than do
black holes.  In softer states, the differences are much more
pronounced.  The neutron stars have two quasi-thermal components --
the disk and the boundary layer -- while black holes have only a disk.
Additionally, the characteristic temperatures of the neutron stars'
disks are higher than those of the black holes because of the
$M^{-1/4}$ scaling of inner disk temperatures.  The combination of these
factors makes the neutron stars have harder spectra in their soft
states than do soft state black holes.

\section{An introduction to our statistical methodology}

We use some methodology for fitting and testing the spectral models
which is non-standard for X-ray spectroscopy, but which have been used
widely in other contexts, and is well-developed.  We use the
\textsc{cstat} option within \textsc{XSPEC 12.8}, following Cash
(1979).  This statistic returns a likelihood function which is
maximized for the best fitting value for a particular model, but does
not yield, in a straightforward manner, a goodness of fit.  We use the
Swift data from 0.5-6.0 keV and the NuSTAR date from 4-20 keV.  {
  These bands are chosen because they are well-calibrated and have
  high ratios of source to background photons for the sources we study
  here.} All source fluxes are reported by taking the unabsorbed model
and integrating between 0.5 and 20 keV.

We then note that the most likely problem with a fit is that the
curvature of the spectral model will be different from the curvature
of the data.  This will lead to maximal differences between cumulative
number of counts in the data and the model, folded through the
response matrix, at the edges of the distribution.  Such a difference
between data and model is identified most readily in an
Anderson-Darling (1954) test.  \textsc{XSPEC} has a routine for
computing the Anderson-Darling parameter as a test statistic, which we
use.  We can then use the Monte Carlo \textsc{goodness} command in
\textsc{XSPEC} to estimate the null hypothesis probability, by running
a set of simulations and determining how often the simulations give
fits with a better Anderson-Darling statistic than the model.  We use
10000 simulations with the \textsc{goodness} command to estimate the
null hypothesis probabilities.

{ We note that we rely on fits to time-integrated spectra for this
  work, rather than examining the source variability.  In principle,
  the source variability could provide very strong constraints on the
  nature of the sources, but the number of counts here is insufficient
  for such an analysis.  E.g., we have approximately 15\% statistical
  uncertainties on hardness ratios between 6-10 keV and 10-20 keV in
  integrations with about 100 ksec of good time, while Smale et
  al. (2003) show that the deviations from the mean in that pair of
  bands is about 25\%; we thus do not have the data quality we need to
  see if the colors follow a Z-track for the sources.}

\section{Individual sources}

\begin{table*}
\begin{tabular}{lllllll}
\hline
Name  & RA &  DEC  &$g$&$u-g$&$g-r$&$r-i$\\
\hline
Bo~153 & 00:43:10.61 &   +41:14:51.4&16.69 &1.81 &0.80 &0.44 \\
Bo~185 & 00:43:37.28 &   +41:14:43.5&16.04 &1.69 &0.77 &0.39\\
Bo~225 & 00:44:29.56 &   +41:21:35.7&14.59 &1.74 &0.77 &0.40 \\
Bo~375 & 00:45:45.56 &   +41:39:42.3&18.04 &1.69 &0.78 &0.35  \\
SKC182C& 00:45:27.32&    +41:32:54.1&19.78 &0.63 &1.27 &0.62\\ 	
\hline
\end{tabular}
\caption{The positions, SDSS magnitudes, and SDSS colors of the clusters, as taken from Peacock et al. (2010).}
\label{clusters}
\end{table*}

For all sources for which we have good NuSTAR spectra, we attempt to
fit three different spectral models: a power law, a disk blackbody,
and a Comptonized blackbody (\textsc{comptt}).  In nearly every case,
we consider absorption with the Galactic value, and absorption which
may float freely; when a statistically acceptable fit is given without
allowing the absorption to float freely, we know already that the model
cannot be rejected, and we do not consider further the variable
absorption case.  Given the low redshift of M31, we do not treat
Galactic absorption and intrinsic absorption as separate components,
but rather treat them as a single component with a summed absorption
column.  We take the Galactic absorption to the M31 fields to be
$10^{21}$ cm$^{-2}$ (Kaberla et al. 2005).  The results of the
different spectral fits are given in Table \ref{fits}.

\begin{table*}
\begin{tabular}{lllllllllll}
\hline
Source & Model & $N_H$ & $\Gamma$&$kT_{in}$&$kT_{e}$ & $R_{in}$& $\tau$ & cstat/dof & AD & Null prob\\
\hline
Bo~153& PLA& $10^{21}$ & 1.7& && & & 639/616& -3.84 & $<10^{-4}$\\
Bo~153& PLB& $10^{21}$ & $1.8\pm0.1$& && & & 618/615& -3.85 & $<10^{-4}$\\
Bo~153& PLC& $4.2^{+1.5}_{-1.0}\times10^{21}$ & $2.1{\pm0.1}$& && & & 567/614& -4.77 &$<10^{-4}$\\
Bo~153&DBBA& $10^{21}$&  &$2.29^{+0.12}_ {-0.11}$& &4&&518/615&-4.86&$<10^{-4}$\\
Bo~153&DBBB& $10^{21}$& &$1.04\pm0.01$&&30&&1510/616&-2.64&$<10^{-4}$\\
Bo~153&DBBC& $\approx 0.0$& & $2.27^{+0.12}_{-0.11}$&&4&&518/614&-5.40&0.004\\
Bo~153&COMPTT&$10^{21}$& & &$2.2\pm0.2$&&$8.6\pm0.8$&506/614&-7.21&0.41\\
\hline
Bo~185& PLA& $10^{21}$ &1.7 && & & & 599/502& -3.39 & $<10^{-4}$\\
Bo~185& PLB& $10^{21}$ & $1.8\pm0.1$&& & & & 593/501& -3.41 & $<10^{-4}$\\
Bo~185& PLC& $7.1^{+1.4}_{-1.2}\times10^{21}$ & $2.2\pm0.1$&& & & & 527/500& -4.79 & 0.0006\\
Bo~185&DBBA& $10^{21}$& &$2.43^{+0.15}_{-0.14}$&&3&&485/502&-5.88&$<10^{-4}$\\
Bo~185&DBBB& $10^{21}$&  &$0.97\pm0.01$&&30&&1559/502&-2.35&$<10^{-4}$\\
Bo~185&DBBC& $\approx 0.0$& & $2.51\pm0.15$&&2&&479/500&-6.58&0.15\\
Bo~185&COMPTT&$10^{21}$& & &$2.1\pm0.2$&&$9.8^{+1.1}_{-1.3}$ &481/500&-6.75&0.80\\
\hline
Bo~225& PLA& $10^{21}$ & 1.7&& & & & 602/603& -4.60 & $<10^{-4}$\\
Bo~225& PLB& $10^{21}$ &  $1.9\pm0.1$&&& & & 556/602& -4.70 & $<10^{-4}$\\
Bo~225& PLC& $3.0^{+0.7}_{-0.6}\times10^{21}$ & $2.1\pm0.1$&& & & & 531/601& -5.46 & 0.0003\\
Bo~225&DBBA& $10^{21}$& &$2.32^{+0.13}_{-0.12}$&&3&&660/602&-3.84&$<10^{-4}$\\
Bo~225&DBBB& $10^{21}$&&$0.97\pm0.01$&&30&&1714/603&-2.32&$<10^{-4}$\\
Bo~225&DBBC& $\approx 0.0$&&$2.40^{+0.13}_{-0.12}$&&3&&614/602&-4.17&$<10^{-4}$\\
Bo~225&COMPTT&$10^{21}$&&&$3.1^{+0.5}_{-0.4}$&&$6.2\pm0.7$&519/601&-7.10&0.31\\
\hline
Bo~375& PLA& $10^{21}$ & 1.7& &&& & 2897/848& -3.35 & $<10^{-4}$\\
Bo~375& PLB& $10^{21}$ & $2.0\pm0.0$&& && & 2273/847& -3.37 & $<10^{-4}$\\
Bo~375& PLC& $5.7^{+0.4}_{-0.2}\times10^{21}$&  $2.5\pm0.0$&& && & 1544/846& -4.30 & $<10^{-4}$\\
Bo~375&DBBA& $10^{21}$&&$1.86\pm0.03$&&13&&937/847&-5.40&$<10^{-4}$\\
Bo~375&DBBB& $10^{21}$&&$1.47\pm0.01$&&30&&1644/848&-4.92 &$<10^{-4}$\\
Bo~375&DBBC& $\approx 0.0$&&$1.93\pm0.03$&&12&&847/846&-6.15&$<10^{-4}$\\
Bo~375&COMPTT&$10^{21}$&&&$1.7\pm0.1$&&$10.3^{+0.4}_{-0.3}$&849/846&-7.15&0.0001\\
Bo~375& DBBPL& $1.5\pm0.7\times10^{21}$ & $2.3\pm0.1$&$1.88^{+0.08}_{-0.07}$ &&12& & 795/844& -8.05 & 0.035\\
Bo~375& DBBCMPT& $10^{21}$ & & $1.54^{+0.31}_{-0.33}$&$2.3^{+2.6}_{-0.4}$&16&$8.4^{+3.7}_{-4.4}$ & 792/844& -8.73 & 0.32\\
\hline
SK182C& PLA& $10^{21}$ & 1.7&& & & &408/441& -7.16 & 0.78\\
SK182C& PLB& $10^{21}$ & $1.7\pm0.1$&&& & & 408/440& -7.02 & 0.69\\
SK182C& PLC& $2.3^{+1.4}_{-1.2}\times10^{21}$ & $1.8\pm0.1$&&& & & 404/439& -6.94 & 0.93\\
SK182C&DBBA& $10^{21}$&&$3.11^{+0.41}_{-0.34}$&&1&&465/440&-3.34&$<10^{-4}$\\
SK182C&DBBB& $10^{21}$&&$0.75\pm0.02$&&30&&997/404&-1.32 &$<10^{-4}$\\
SK182C&DBBC& $\approx 0.0$&&$3.24^{+0.43}_{-0.36}$&&1&&452/439&-3.55&$<10^{-4}$\\
SK182C&COMPTT&$10^{21}$&&&$45.1^{+1839.7}_{-45.1}$&&$1.1^{+4.1}_{-1.1}$&408/439&-7.06&0.92\\
\hline
\end{tabular}
\caption{The table of spectral fits for the five sources.  The first
  column gives the source names.  The second column gives the
  different models used: PLA -- power law with frozen $N_H$ and frozen
  $\Gamma$; PLB -- power law with frozen $N_H$, but free $\Gamma$; PLC
  -- power law with both parameters free; DBBA -- disk blackbody
  model with frozen $N_H$ but free normalization and temperature; DBBB
  -- disk blackbody with frozen $N_H$, frozen normalization to 0.1
  (consistent with 30 km inner radius at M31 distance for a face-on
  disk) and free temperature; DBBC -- disk blackbody with all
  parameters free; COMPTT, where only a single model is fitted, which
  has frozen $N_H$, and seed photon temperature frozen to 0.1 keV;
  DBBPL -- disk blackbody plus power law with all parameters free; 
  DBBCMPT - disk blackbody plus COMPTT with  
  frozen $N_H$ and frozon seed photon temperature of 0.1 keV.
  The third column gives the $N_H$ used for the fit in cm$^{-2}$.  The fourth
  column gives $\Gamma$, the spectral index for the power law spectra,
  defined such that the differential number of photons as a function
  of energy, $\frac{dN}{dE}$ scales as $E^{-\Gamma}$.  The fifth column gives the inner
  disk temperature from the \textsc{diskbb} model. The sixth column gives the electron
  temperature in the corona for the \textsc{comptt} model.  The seventh
  column gives the inner disk radius in km for the \textsc{diskbb} model,
  assuming a face-on disk and no colour correction.  The eighth column
  gives the optical depth of the thermal comptonization model.  The
  nineth column gives the value of the Cash statistic and the number
  of degrees of freedom for the fit.  The tenth column gives the
  logarithm of the Anderson-Darling statistic.  The eleventh column gives
  the fraction of the simulations made using \textsc{goodness} that
  were statistically as bad as the model fit.  Where no simulations
  were as bad as the model fit, $<10^{-4}$ is placed in this column.}

\label{fits}
\end{table*}

\subsection{Bo~153}

\begin{figure}
\includegraphics[width=2.5 in,angle=-90]{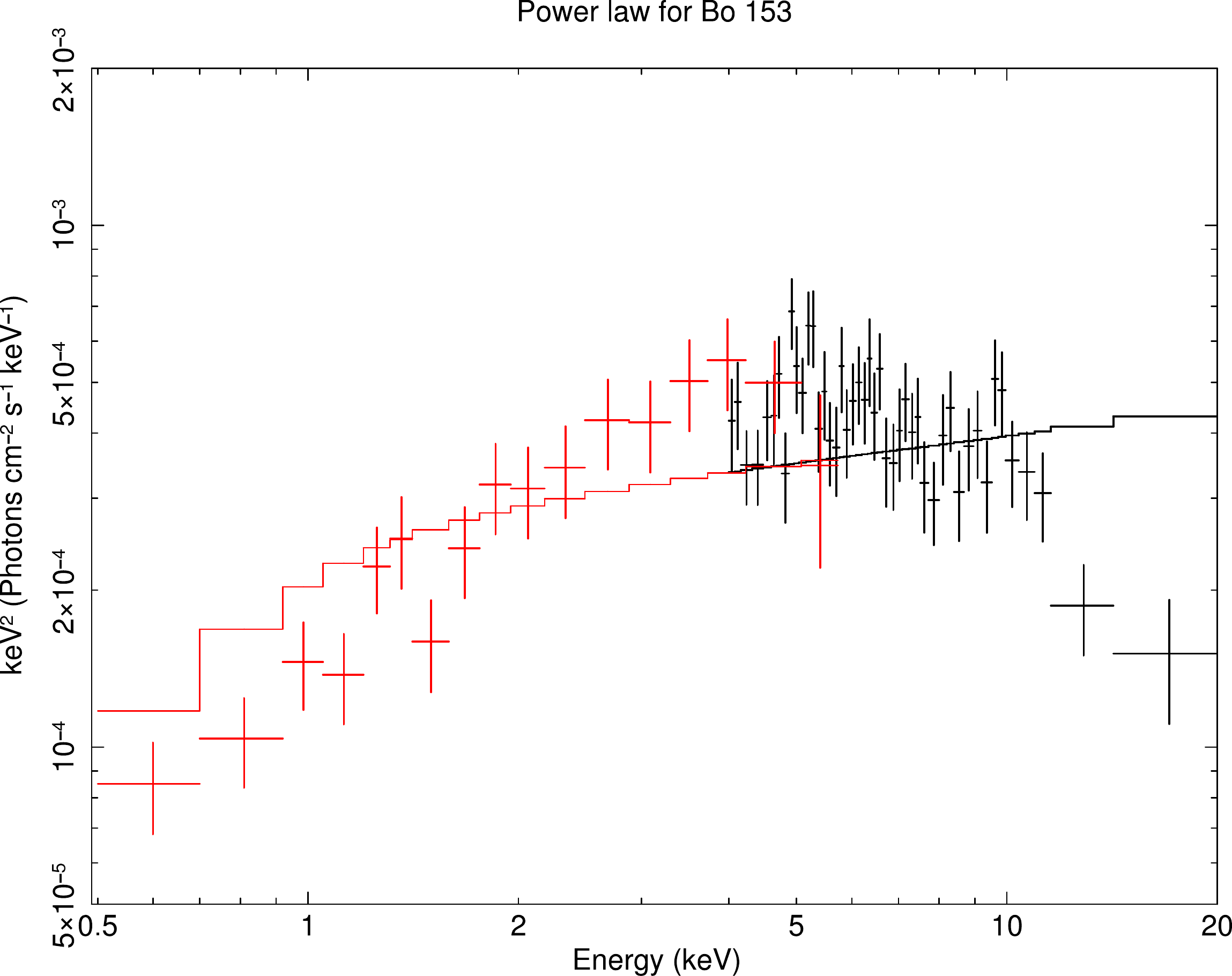}
\caption{The best fitting power law model with absorption frozen to
  $10^{21}$ cm$^{-2}$ for Bo~153's X-ray spectrum.  The data are
  plotted after rebinning either until a signal to noise
  of 5 is reached, or 100 bins have been used, but the input
  spectra grouped to one count per bin have been used.  From the plot,
  it is clear that the data have a greater level of curvature than the
  model does.  The Swift data, and the model convolved through the
  Swift response function are in red while the NuSTAR data and the
  model convolved through NuSTAR's response function are in black.}
\label{powerlaw}
\end{figure}

\begin{figure}
\includegraphics[width=2.5 in,angle=-90]{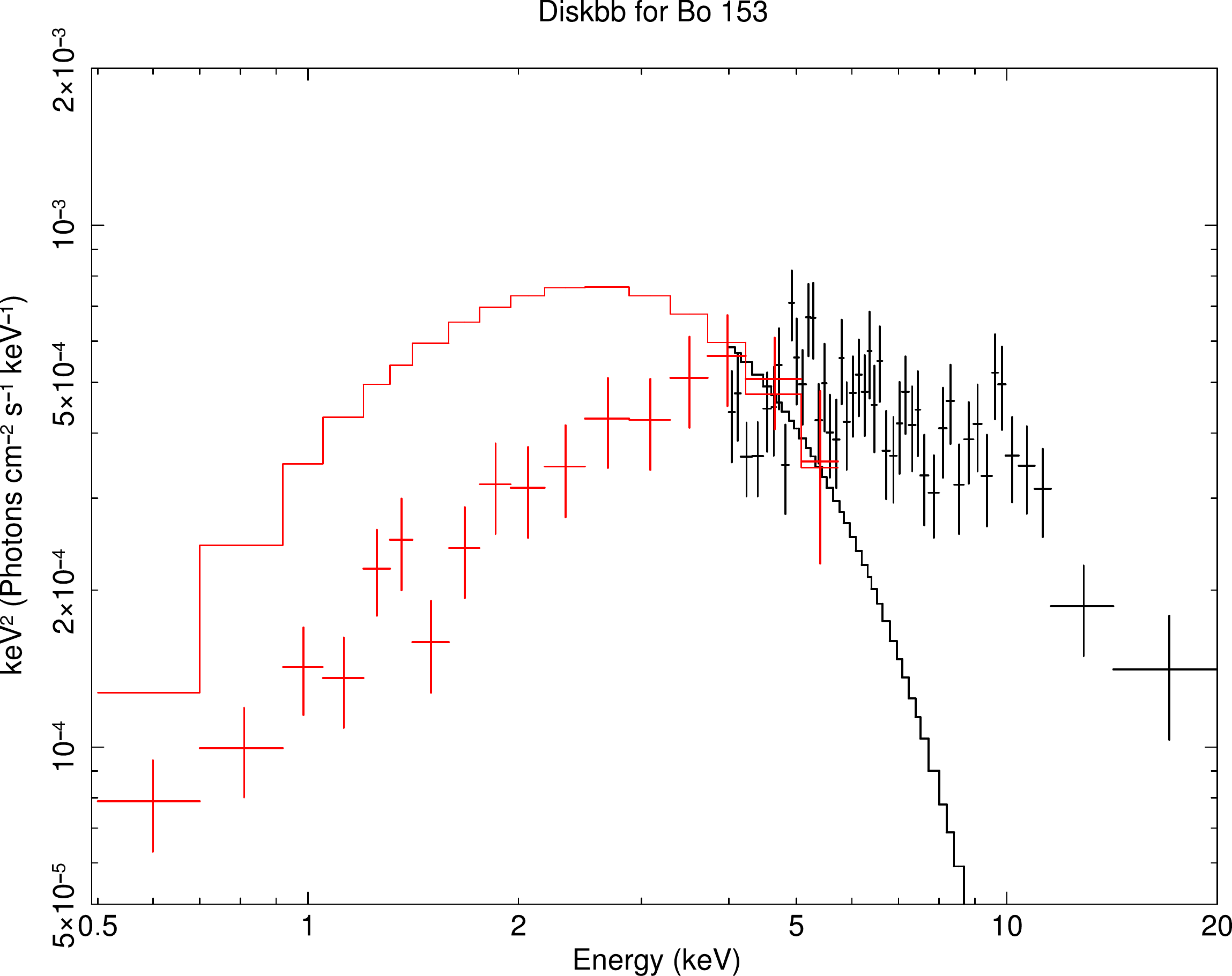}
\caption{The best fitting disk blackbody model with absorption frozen
  to $10^{21}$ cm$^{-2}$ for Bo~153's X-ray spectrum, and the
  normalization frozen to a physically plausible value.  The data are
  plotted after rebinning either until a signal to noise
  of 5 is reached, or 100 bins have been used, but the input
  spectra grouped to one count per bin have been used. That the real
  spectrum is harder than any reasonable disk model is obvious from
  the plot.  The Swift data, and the model convolved through the Swift
  response function are in red while the NuSTAR data and the model
  convolved through NuSTAR's response function are in black.}
\label{diskbb}
\end{figure}

\begin{figure}
\includegraphics[width=2.5 in,angle=-90]{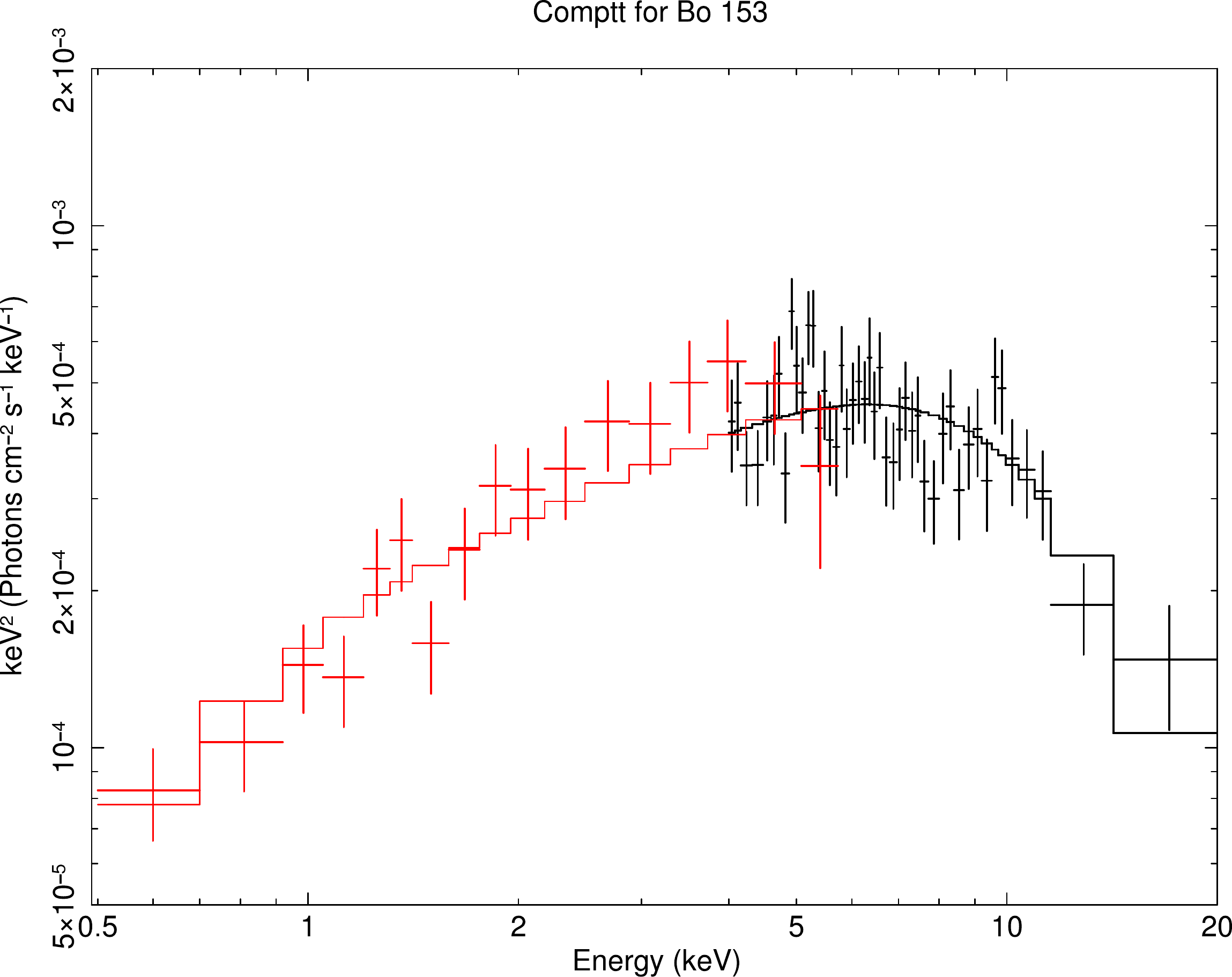}
\caption{The best fitting thermal Comptonization model with absorption
  frozen to $10^{21}$ cm$^{-2}$ for Bo~153's X-ray spectrum.  The data
  are plotted after rebinning either until a signal to
  noise of 5 is reached, or 100 bins have been used, but the input
  spectra grouped to one count per bin have been used. The model can
  be seen to be a good description of the data.  The Swift data, and
  the model convolved through the Swift response function are in red
  while the NuSTAR data and the model convolved through NuSTAR's
  response function are in black.  It is clear that the model not only
  provides a good statistical fit to the data, but also matches the
  curvature of the data.}
\label{comptt}
\end{figure}

Bo~153 was suggested by Barnard et al. (2011) to be a strong candidate
for being a globular cluster black hole on the basis of appearing to
fit well to a low/hard state spectrum while being at a luminosity
(varying in the range from 0.8--2.4$\times10^{38}$erg/sec {
  in the 0.3--10.0 keV band}) above which neutron stars do not show
low hard states. { With NuSTAR, the single power law model
  gives an unacceptable fit -- clearly the data show more curvature
  than a single power law allows.}
disk blackbody model with free temperature and { column
  density } produces a fit which is marginally statistically
acceptable, but which requires an unphysically small inner disk radius
(i.e. much less than a Schwarzschild radius), unphysically large
temperatures (i.e. $>$ 2 keV), { and no Galactic column
  density}.  When the model is forced to have an inner disc radius of
30~km, the fit is no longer statistically acceptable.  The
\textsc{comptt} model provides a fit which is statistically
acceptable, and which has parameters in line with typical Z sources
and typical bright atoll sources.  The flux from the \textsc{comptt}
model, correcting for absorption, is { $1.7\times10^{-12}$
  erg/sec/cm$^{2}$, corresponding to a luminosity of
  $1.2\times10^{38}$ erg/sec} for a distance of 784 kpc (Stanek \&
Garnavich 1998), which is also a typical value for a Z source.
Figures are presented for the power law fit -- Figure \ref{powerlaw},
the disk blackbody fit -- Figure \ref{diskbb}, and the thermal
Comptonization model fit -- Figure \ref{comptt}.  These figures are
representative of the results for all the sources, so we do not
present figures for the fits to the other sources.

\subsection{Bo 185}

The results for this source are quite similar to those for Bo~153.
This source was also claimed by Barnard et al. (2011) to be a strong
globular cluster black hole candidate on the ground of being a bright
hard state object.  Like for Bo~153, we find that the power law model
fits are not statistically acceptable, and that the statistically
acceptable disk blackbody model fits have unphysically small inner
disk radii and unphysically large temperatures.  We find that the
thermal comptonization model gives a fit that is typical of Z-sources
and bright atoll sources.  The flux from the model, correcting for
absorption, is { $1.0\times10^{-12}$ erg/sec/cm$^{-2}$,
  corresponding to a luminosity of $7.3\times10^{37}$ erg/sec}, which
is, again, typical for Z-sources and bright atoll sources.

\subsection{Bo 225}
An additional bright X-ray source in M31 is Bo~225.  This object is
less well studied than the two previously discussed sources and has
not been claimed in the past to be a black hole candidate.  The only
model that fits this source well is the \textsc{comptt} model.
this source, then, we may have run into the limitations of using such
a simple thermal Comptonization model, and the data may be justifying
a slightly high level of complexity.  Nonetheless, this model is
clearly the best of the group we have tried.  The flux from the
unabsorbed \textsc{comptt} model is
$1.3\times10^{-12}$erg/sec/cm$^{-2}$, corresponding to a luminosity of
$9\times10^{37}$ erg/sec.  The combination of the luminosity of the
source, and the fact that \textsc{comptt} provides both a
statistically acceptable fit and reasonable parameter values,
indicates that the source can be confidently identified as a Z-source
or a bright atoll source.

{
\subsection{Bo~375}

Bo~375 has also been observed with NuSTAR and Swift.  This is a bright
source which has been previously classified as a neutron star (Barnard
et al 2008).  For this source, none of the models with a single
continuum component provides a good fit to the data.  The data can be
well fit with a model consisting of a disk blackbody plus a
comptonized blackbody, which is one of the models often used to fit
Z-sources and bright atoll sources (null hypothesis probability of
0.31) and marginally well-fit by a disk blackbody plus power law model
(null hypothesis probability of 0.03).  The inner disk radius for the
disk blackbody plus power law model is unphysically small (12 km), so
this model is additionally disfavored.  The flux from the disk
blackbody plus comptonized blackbody model, correcting for absorption,
is { $8.1\times10^{-12}$erg/sec/cm$^{-2}$, corresponding to
  a luminosity of $6\times10^{38}$ erg/sec} for a distance of 784 kpc.
This value is slightly above the Eddington luminosity for a
1.4$M_\odot$ neutron star.  This value is slightly above the highest
luminosity seen from Sco~X-1 of $4.5\times10^{38}$ erg/sec (Barnard et
al. 2003), which is robust given the geometric parallax distance
(Bradshaw et al. 1999), but the discrepancy can be explained if the
neutron star in Bo~375 is a bit more massive than the neutron star in
Sco~X-1, or if the neutron star in Bo~375 is accreting hydrogen-poor
gas.  Given that ultracompact X-ray binaries represent a substantial
fraction of the X-ray binaries in Milky Way globular clusters (Stella
et al. 1987; Dieball et al. 2005; Zurek et al. 2009), this latter
interpretation would not be surprising.

Barnard et al. (2008) had previously found with XMM-Newton data that a
single power law could not fit the spectrum of that source, suggesting
that it is a neutron star.  We thus favor a neutron star
interpretation for the data, as it is consistent with both our
analysis and that of Barnard et al. (2008).

\subsection{SK182C}

SK182C is located in the same field of view as Bo~375, and hence is
included in the same observations..  This source is fainter than the
others, so our ability to rule out models is somewhat diminished.  For
this source, we find both the power law and \textsc{comptt} models to
be statistically acceptable and to have reasonable parameter values.

The flux from the model, correcting for absorption, is
$7.6\times10^{-13}$erg/sec/cm$^{-2}$, corresponding to a luminosity of
$5\times10^{37}$ erg/sec at the distance to M31.  For the spectra
we analyze here, we find that either a low hard state black hole model
or a neutron star model could fit well to the data.  The disk
blackbody models are all statistically unacceptable.  Given that the
source is at about 5\% of the Eddington luminosity for an 8 $M_\odot$
black hole, and that black holes at such a luminosity are usually in
soft states unless they are caught in the rise of a transient outburst
(Maccarone 2003; Kalemci et al. 2013), the black hole interpretation
is disfavored for this source, but not as strongly as for the other
sources in the sample.  This source also has no previous
identification as either a black hole or a neutron star.}

\section{Discussion}

These results cast doubt on many of the other claims of globular
cluster black holes in M31 globular clusters.  Many of these are based
on the same methodology as the claims for Bo~153 and Bo~185.  The
results also illustrate the importance of having a broad bandpass, as
from NuSTAR, for making classifications of black hole and neutron star
spectra.  In particular, the Z-source/bright atoll source spectral
models can be seen to be relatively similar to power law spectra, as
long as most of the counts are obtained below about 10 keV where the
spectra break sharply.  This is, notably, where the responses of
Chandra and XMM-Newton start to become poor.

{ Some additional support for the neutron star nature of the
  sources can come from looking at the Milky Way's population of
  persistent black hole candidates.  The only dynamically confirmed
  black hole candidate which is persistent is Cygnus X-1 (Gies \&
  Bolton 1986; Caballero-Nieves et al. 2009; Orosz et al. 2011), which
  spends most of its time in a hard state, but which is likely to be
  immune from the full ionization instability due to having a high
  mass donor star, and being wind-fed so that the circularization
  radius of the disc is smaller than for Roche lobe overflow from a
  star with the same orbital period (see discussion in Smith et
  al. 2002a).  Cygnus X-3 represents a similar case, although its
  dynamical confirmation is not clear (Szostek \& Zdziarski 2008).
  SS~433 is even less securely a black hole, and is likely
  intrinsically super-Eddington but observed edge-on so that only
  scattered X-rays are seen (e.g. Charles et al. 2004).  4U~1957+11 is
  not dynamically confirmed, but appears to spend most of its time in
  soft states (Gomez et al. 2015).  

There are two other persistent sources, 1E~1740.7-2942 and
GRS~1758-258, which appear to be long X-ray periodicities (12.73 and
18.45 days, respectively), and which spend significant fractions of
their time in hard states, but these objects are not dynamically
confirmed black holes, they have unknown donor types (Smith et
al. 2002b), and the periods are significantly longer than expected for
tidal capture products. The persistent hard state black hole sources
with low mass donors thus may not exist at all, and are clearly, at
most, a small fraction of the total source population bright enough to
be detected with all-sky instruments in the Milky Way (which is a
similar luminosity limit to the luminosity needed to detect a source
at all in M31).  There are, of course, a number of relatively steady
quiescent black hole sources, which might be regarded as persistent
hard state sources, but all of these which are dynamically confirmed
as black holes have undergone large outbursts.  

There are selection effects against dynamical confirmation of black
holes in persistently bright X-ray binaries.  Nonetheless, there is
only one candidate persistent black hole low mass X-ray binary whose
orbital period is short enough to be in the $\sim 10$hrrange where
tidal capture might work, 4U 1957+11.  There are many black hole X-ray
binaries with both shorter and longer orbital periods which are
transients, and there are no other persistent X-ray emitters whose
X-ray emission properties mark them as likely black hole accretors.
As a result, it seems highly unlikely that black hole X-ray binaries
that form from tidal capture are typically persistent, but it cannot
be excluded that this might occasionally happen.

}

\subsection{Disk winds and the transient problem}

{ 

One of the original motivations for considering these objects to be
persistent black hole binaries formed by tidal capture was that such
objects had been predicted to exist. Repeated claims exist in the
literature (e.g., Kalogera et al. 2004; Barnard et al. 2009) that
tidal capture black hole X-ray binaries should be persistently X-ray
bright. Thus, our finding that these objects in M31 are likely neutron
stars gives us good cause to re-consider some of the assumptions that
went into these claims.  In particular, it is worth considering why it
may actually be unlikely for a large population of persistent black
hole X-ray binaries to exist, especially in the orbital period range
expected for tidal capture products.

Standard binary evolution and disk instability theory predicts that
systems with orbital periods of about 10 hours, the range expected
from tidal captures, should be persistent sources (King et al. 1996;
Kalogera et al. 2004). However, as we describe below, this claim is
not in line with phenomenology of outburst behaviour from black hole
X-ray binaries, and we suggest that mass loss in disk winds can
explain the discrepancy between observation and theory.

}

Comparison of figure 1 of King et al. (1996), which shows the expected
mass transfer rates as a function of orbital period, and figure 7 of
Lasota (2001), which shows observed mean accretion rates as a function
of orbital period indicates a clear discrepancy between the two
values.  King et al. (1996) find that the typical mass transfer rates
in black hole X-ray binaries in that orbital period range should be
$10^{-10}-10^{-9} M_\odot$/yr, while the recurrence times of X-ray
transients indicate that the mass transfer rates are more typically a
few times $10^{-11} M_\odot$/yr.  Indeed, the discrepancy is already
implicitly noted by King et al. (1996) themselves, who point out that
the known low mass X-ray binaries with black hole primaries are
predominantly soft X-ray transients, rather than persistent emitters.
Our finding that these persistent sources in M31 are more likely
neutron stars than black holes underscores this point. 

A few possible explanations exist { for the paucity of persistent
  black hole emitters.  Perhaps the simplest is to invoke a mechanism
  that lowers } the accretion rate onto the cental black hole {
  relative to that} predicted from the prescriptions for binary
evolution used in King et al. (1996).  { Two possible ways to
  change the mass accretion rates are to change the mass transfer
  rates by invoking alternative prescriptions for magnetic braking,
  and invoking non-conservative mass transfer due to disk winds.}  The
now-good agreement between theory and data for models of evolution of
cataclysmic variables (Knigge et al. 2011) casts doubt on the {
  possibility that magnetic braking prescriptions are badly flawed},
unless the donor stars in black hole X-ray binaries are considerably
more bloated than those in cataclysmic variables.

{ Disk winds, on the other hand, show clear evidence of being
  present and important in X-ray binaries.}  If we take, { for
  example}, the best studied quiescent X-ray binary, A0620-00 we see
that it has an orbital period of 7.1 hours, and a donor mass of 0.4
$M_\odot$, meaning that its donor mass is about 0.6 times as large as
that of an unevolved star filling its Roche lobe, given the well known
relationship between donor star mass and orbital period for main
sequence stars.  This produces a reduction in mass transfer rate by a
factor of about 3.5 -- a substantial factor, but not one large enough
to explain the discrepancy between the observed mean outburst fluence
averaged over the best estimate of the source duty cycle and the
predicted mass transfer rates.

Therefore, it appears more likely that disk winds during the outbursts
of black hole X-ray binaries lead to highly non-conservative mass
transfer { and hence make the mean accretion rate by the compact
  object less than the mass loss rate by the donor star}.  Three
methods have been used to estimate the mass loss due to disk winds,
and all result in the finding that $\sim90\%$ of the mass lost by the
donor star is also lost by the accretion disk.  One method for making
the estimate is the result above, that transient outbursts seem to
have recurrence timescales about 10 times too long, and that some
objects which would be expected to be persistent are, in fact,
transient.  Another is that estimates can be made of the depths of
absorption lines seen from the accretion disks, the opening angles of
the disk winds, and the chemical composition and ionization state of
the absorbing gas, and convert these to mass loss rates in the disk
wind (e.g. Neilsen et al. 2011).  Additionally, one can estimate the
mass transfer rate from the luminosity of the hot spot where the
accretion stream impacts the outer accretion disk, and compare with
the quiescent X-ray luminosity and with the outburst duty cycle
(e.g. Froning et al. 2011).

Finally, two short-period X-ray binaries have period derivatives that
cannot be well explained in the light of standard binary evolution
scenarioes (Gonzalez Hernandez et al. 2014).  If mass loss is the
cause of the latter effect, then strong mass loss takes place even for
quiescent X-ray binaries.  This is at odds with prominent
interpretations of recent findings that strong X-ray absorption lines
are seen only when sources are in X-ray soft states (Neilsen \& Lee
2009; Ponti et al. 2012).  On the other hand, if the ionization state
of the wind, rather than the presence of the wind, is what changes at
state transitions, then the claims can be reconciled.  Additionally,
there are detections of single-peaked emission lines in GX~339-4 in
hard states (Wu et al. 2001), and single-peaked lines from accretors
are often associated with scatter broadening of lines by disk winds
(e.g. Shlosman \& Vitello 1993; Knigge \& Drew 1996 Murray \& Chiang
1996; Sim et al. 2010).

{ All these methods of estimating the actual accretion rates in
  black hole X-ray binaries, and the amount of mass loss in their
  winds,} suggest that mass transfer is in black hole binaries is {
  highly} non-conservative.  { Systems which would be persistent
  can then be forced into quiescent regimes because the mass loss will
  reduce the central accretion rates and reduce the irradiation of the
  outer disks by the inner disks; additionally, just the mere loss of
  mass will reduce the gas density, and hence the gas temperature in
  the outer disks, even in the absence of irradiation.}  The duty
cycles of bright sources are then suppressed compared to what would be
expected. Importantly, in cataclysmic variables, the mass loss rates
in the wind are very small compared to mass accretion rates by the
central white dwarfs (Vitello \& Shlosman 1988; Knigge et al. 1995),
so disk winds should not be important for CV evolution. The same may
be true for neutron star accretors.  Thus, in hindsight, one should
not expect a substantial population of persistent black hole X-ray
binaries, even if the systems are formed by tidal capture, and
persistent sources should be expected to more commonly be neutron
stars, as we have found here.

At the same time, a more detailed treatment of disk winds would be
well-justified.  In this paper, we have discussed disk winds only in
the context of removing mass from the accretion disk at a constant
accretion rate.  A more detailed treatment of nonconservative mass
transfer would also consider the angular momentum carried away by the
mass lost from the accretion disk and its subsequent effects on the
mass transfer rate itself.  Such a treatment lies beyond the scope of
this paper, but would be well-justified in light of recent
developments.

{ \section{Summary}

We have examined 5 globular cluster sources at $L_X \sim 10^{38}$
erg/sec in M31 with a combination of data from Swift and NuSTAR.  We
have found that in all cases, the data can be well fitted with high
optical depth ($\tau$$\sim$$3-10$), low temperature ($k_{B}T$$<{10}$)
keV comptonization models, while for four of the five sources, the
data argue strongly against a single power law model fitting the data,
and for all the sources, the data argue against a multi-temperature
blackbody disk model fitting the data.  As a result, we argue that
these sources are all likely to be high accretion rate neutron star
X-ray binaries, although the data are not presently good enough to
determine whether they are ``Z-sources'' or bright atoll sources in
soft states.  Bolstering the idea that these are neutron stars is the
dearth of black hole X-ray binaries.  Because past theoretical work
has argued that the accretion rates expected within a particular range
of orbital periods should yield persistent black hole X-ray binaries,
we have discussed how non-conservative mass transfer due to disc winds
leads to a violation of the assumptions of the past work.  }

\section{Acknowledgments}

TJM thanks Christian Knigge for an illuminating talk at ``The Physics
of Cataclysmic and Compact Binaries'', and Helena Uthas, Joe
Patterson, Christian Knigge and Jeno Sokoloski for having organized
the meeting.  He also thanks Joey Neilsen, Chris Done, and Maria Diaz
Trigo for useful discussions about disk winds.  AZ acknowledges
funding from the European Research Council under the European Union's
Seventh Framework Programme (FP/2007-2103)/ ERC Grant Agreement
n. 617001.

\label{lastpage}


\begin{thebibliography}{99}
\bibitem{}Anderson T.W., Darling D.A., 1952, Annals of Mathematical Statistics, 23, 193
\bibitem{}Bachetti M., et al., 2014, Nature, 514, 202
\bibitem{}Barnard R., Church M.J., Ba\/luci\'nska-Church M., 2003, A\&A, 405, 237
\bibitem{}Barnard R., et al., 2008, ApJ, 689, 1215
\bibitem{}Barnard R., Garcia M., Li Z., Primini F., Murray S.S., 2011, ApJ, 734, 79
\bibitem{}Barret D., McClintock J.E., Grindlay J.E., 1996, ApJ, 473, 963
\bibitem{}Basu-Zych A.R., et al., 2013, ApJ, 774, 152
\bibitem{}Baumgardt H., Hut P., Makino J., McMillan S., Portegies Zwart S., 2003, ApJ, 582L, 21
\bibitem{}Binder B., Gross J., Williams B.F., Simons D., 2015, arxiv/1506.03092, accepted to MNRAS
\bibitem{}Bradshaw C.F. Fomalant E.B., Geldzahler B.J., 1999, ApJ, 512L, 121
\bibitem{}Brassington N.J., et al., 2008, ApJS, 179, 142 
\bibitem{}Brorby M., Kaaret P., Prestwich A., 2014, MNRAS, 441, 2346
\bibitem{}Caballero-Nieves S., et al., 2009, ApJ, 701, 1895
\bibitem{}Cannizzo J.K., Chen W., Livio M., 1995, ApJ, 454, 880 
\bibitem{}Cannizzo J.K., Wheeler J.C., Ghosh P., 1985, ASSL, 113, 307
\bibitem{}Casares J., Jonker P.G., 2014, SSRv, 183, 223
\bibitem{}Cash W., 1979, ApJ, 228, 939
\bibitem{}Charles P.A., et al., 2004, RMxAC, 20, 50
\bibitem{}Chomiuk L., Strader J., Maccarone T.J., Miller-Jones J.C.A., Heinke C., Noyola E., Seth A.C., Ransom S., 2013, ApJ, 777, 69
\bibitem{}Church M.J., Baluci\'nska-Church M., 1995, A\&A, 300, 441
\bibitem{}Clark G.W., 1975, ApJ, 195L, 143
\bibitem{}Crowther P., Barnard R., Carpano S., Clark J.S., Dhillon V.S., Pollock A.M.T., 2010, MNRAS, 403, 41
\bibitem{}Davis S.W., Blaes O.M., Hubeny I., Turner N.J., 2005, ApJ, 621, 372
\bibitem{}Dieball A., Knigge C., Zurek D.R., Shara M.M., Long K.S., Charles P.A., Hannikainen D.C., van Zyl L., 2005, ApJ, 634L, 105
\bibitem{}Dunn R.J.H., Fender R.P., K\"ording E.G., Belloni T., Cabanac C., 2010l, MNRAS, 403, 61
\bibitem{}Emparan R., Fabbri A., Kaloper N., 2002, JHEP, 08, 043
\bibitem{}Fabian A.C., Pringle J.E., Rees M.J., 1975, MNRAS, 172P, 15
\bibitem{}Fan Z., Ma J., de Grijs R., Zhou X., 2008, MNRAS, 385, 1973
\bibitem{}Farr W.,M., Sravan N., Cantrell A., Kreidberg L., Bailyn C.D., Mandel I., Kalogera V., 2011, ApJ, 741, 103
\bibitem{}Ferrarese L., Merritt D., 2000, ApJ, 539L, 9
\bibitem{}Froning C.S., et al., 2011, ApJ, 743, 26
\bibitem{}Fryer C.L., Kalogera V., 2001, ApJ, 554, 548
\bibitem{}Galetti S., Federici L., Bellazzini M., Fusi Pecci F., Macrina S., 2004, A\&A, 416, 917
\bibitem{}Gebhardt K., et al., 2000, ApJ, 539L, 13
\bibitem{}Giacconi R., Gursky H., Kellogg E., Schreier E., Tananbaum H., 1971, ApJ, 167L, 67
\bibitem{}Gierli\'nski M., Done C., 2002, MNRAS, 337, 1373
\bibitem{}Gies D.R. \& Bolton C.T., 1986, ApJ, 304, 371
\bibitem{}Gnedin O.Y., Maccarone T.J., Psaltis D., Zepf S.E., 2009, ApJ, 705, 168L
\bibitem{}Gomez S., Mason P.A., Robinson E.L., 2015, arxiv/1506.00181 
\bibitem{}Gonzalez Hernandez J.I., Rebolo R., Casares J., 2014, MNRASL, 438, L21
\bibitem{}Grindlay J.E., Gursky H., Schnopper H., Parsignault D.R., Heise J., Brinkman A.C., Schrijver J., 1976, ApJ, 205L, 127
\bibitem{}Haggard D., Cool A.M., Heinke C.O., van der Marel R., Cohn H.N., Lugger P.M., Anderson J., 2013, ApJ, 773L, 31 
\bibitem{}Hasinger G., van der Klis M., 1989, A\&A, 225, 79
\bibitem{}Heggie D.C., Giersz M., 2014, MNRAS, 439, 2459
\bibitem{}Hills J.G., 1976, MNRAS, 175P, 1
\bibitem{}Homan J., et al., 2001, ApJS, 132, 377
\bibitem{}Kalberla P.M.W., Burton W.B., Hartmann D., Arnal E.M., Bajaja E., Morras R. Poppel W.G.L., 2005, A\&A, 440, 775
\bibitem{}Kalemci E., Dincer T., Tomsick J.A., Buxton M.M., Bailyn C.D., Chun Y.Y., 2013, 779, 95
\bibitem{}Kalogera V., Baym G., 1996, ApJ, 470L, 61
\bibitem{}Kalogera V., King A.R., Rasio F.A., 2004, ApJ, 601L, 171
\bibitem{}Knigge C., Baraffe I., Patterson J., 2011, ApJS, 194, 28 
\bibitem{}King A.R., Kolb U., Burderi L., 1996, ApJ, 464L, 127 
\bibitem{}Knigge C., Woods J.A., Drew J.E., 1995, NRAS, 273, 225
\bibitem{}Knigge C., Drew J.E., 1996, MNRAS, 281, 1352
\bibitem{}Kulkarni S., Hut P., Mcmillan S., 1993, Nature, 364, 421
\bibitem{}Lasota J-P., 2001, NewAR, 45, 449
\bibitem{}Laycock S.G.T., Cappallo R.C., Moro M.L., 2015a, MNRAS, 446, 1399
\bibitem{}Laycock S.G.T., Maccarone T.J., Christodoulou D.M., 2015, 452L, 31
\bibitem{}Lin D., Remillard R.A., Homan J., 2009, ApJ, 696, 1257
\bibitem{}Linden T., Kalogera V., Sepinsky J.F., Prestwich A., Zezas A., Gallagher J.S., 2010, ApJ, 725, 1984
\bibitem{}Maccarone T.J., 2003, A\&A, 409, 697
\bibitem{}Maccarone T.J., Coppi P.S., 2003, MNRAS, 338, 189
\bibitem{}Maccarone T.J., 2004, MNRAS, 351, 1049
\bibitem{}Maccarone T.J., Kundu A., Zepf S.E., Rhode K.L., 2007, Nature, 445, 183
\bibitem{}Maccarone T.J., Kundu A., Zepf S.E., Rhode K.L., 2011, MNRAS, 410, 1655
\bibitem{}Mapelli M., Ripamonti E., Zampieri L., Colpi M., Bressan A., 2010, MNRAS, 408, 234
\bibitem{}Maraschi L., Cavaliere A., 1977, Highlights of Astronomy, 4-1, 127
\bibitem{}McClintock J.E., Remillard R.A., 1986, ApJ, 308, 110
\bibitem{}Miller M.C., Hamilton D.P., 2002, MNRAS, 330, 232
\bibitem{}Mitsuda K., et al., 1984, PASJ, 36, 741
\bibitem{}Mitsuda K., Inoue H., Makamura N., Tanaka Y., 1989, PASJ, 41, 97
\bibitem{}Miyamoto S., Kimura K., Kitamoto S., Dotani T., Ebisawa K.,1991, ApJ, 383, 784
\bibitem{}Miyamoto S., Kitamoto S., Hayashida K., Egoshi W., 1995, ApJ, 442L, 13
\bibitem{}Morscher M., Parrabiraman B., Rodriguez C., Rasio F.A., umbreit S., 2015, ApJ, 800, 9
\bibitem{}Murray N., Chiang J., 1996, Nature, 382, 789
\bibitem{}Neilsen J., Lee J.C., 2009, Nature, 458, 481
\bibitem{}Neilsen J., Remillard R.A., Lee J.C., 2011, ApJ, 737, 69
\bibitem{}Noyola E., Gebhardt K., Bergmann M., 2008, ApJ, 676, 1008
\bibitem{}Olive J.F., Barret D., Boirin L., Grindlay J.E., Swank J.H., Smale A.P., 1998, A\&A, 333, 942
\bibitem{}Orosz J.A., McClintock J.E., Aufdenberg J.P., Remillard R.A., Reid M.J., Narayan R., Gou L., 2011, ApJ, 742, 84
\bibitem{}\"Ozel F., Psaltis D., Narayan R., McClintock J.E., 2010, ApJ, 725, 1918
\bibitem{}Peacock M.B., Maccarone T.J., Knigge C., Kundu A., Waters C.Z., Zepf S.E., Zurek D.R., 2010, MNRAS, 402, 803
\bibitem{}Pecaut M.J., Mamajek E.E., 2013, ApJS, 208, 9
\bibitem{}Ponti G., Fender R.P., Begelman M.C., Dunn  R.J.H., Neilsen J., Coriat M., 2012, MNRAS, 422L, 11
\bibitem{}Prestwich A., et al., 2007, ApJ, 669, L21
\bibitem{}Psaltis D., 2007, PhRvL, 98, 1101
\bibitem{}Redmount I.H., Rees M.J., 1989, ComAp, 14, 165
\bibitem{}Reid M.J., McClintock J.E., Steiner J.F., Steeghs D., Remillard R.A., Dhawan V., Narayan R., 2014, ApJ, 796, 2 
\bibitem{}Remillard R.A., Lin D., Cooper R.L., Narayan R., 2006, ApJ, 646, 407
\bibitem{}Shafee R., McClintock J.E., Narayan R., David S.W., Li L.,  Remillard R.A., 2006, ApJ, 636, L113
\bibitem{}Shakura N.I., Sunyaev R.A., 1973, A\&A, 24, 337
\bibitem{}Shimura T., Takahara F., 1995, ApJ, 445, 780
\bibitem{}Shlosman I., Vitello P., 1993, ApJ, 409, 372
\bibitem{}Sigurdsson S., Hernquist L., 1993, Nature, 364, 423
\bibitem{}Sim S.A., Proga D., Miller L., Long K.S., Turner T.J., 2010, MNRAS, 408, 1396
\bibitem{}Sippel A.C., Hurley J.R., 2013, MNRAS, 430L, 30
\bibitem{}Smale A.P., Homan J., Kuulkers E., 2003, ApJ, 590, 1035
\bibitem{}Smith D.M., Heindl W.A., Swank J.H., 2002a, ApJ, 569, 362
\bibitem{}Smith D.M., Heindl W.A., Swank J.H., 2002b, ApJ, 578L, 129
\bibitem{}Stanek K.Z., Garnavich P.M., 1998, ApJL, 503, 131 
\bibitem{}Stella L., Priedhorsky W., White N.E., 1987, ApJ, 312L, 17
\bibitem{}Strader J. Chomiuk L., Maccarone T.J., Miller-Jones J.C.A., Seth A.C., 2012a, Nature, 490, 71
\bibitem{}Strader J., Chomiuk L., Maccarone T.J., Miller-Jones J.C.A., Seth A.C., Heinke C.O., Sivakoff G.R., 2012b, ApJ, 750L, 27
\bibitem{}Szsostek A., Zdziarski A.A., 2008, MNRAS, 386, 593
\bibitem{}Tananbaum H., Gursky H., Kellogg E., Giacconi R., Jones C., 1972, ApJ, 177L, 5
\bibitem{}Thorne K.S., \& Price R.H., 1975, ApJ, 195L, 101
\bibitem{}Titarchuk L., 1994, ApJ, 434, 570
\bibitem{}Truss M., Done C., 2006, MNRAS, 368L, 25
\bibitem{}van den Heuvel E., 1983, in W.Lewin, E. van den Heuvel (eds), {\it Accretion Driven Stellar X-ray Sources}, Cambridge University Press: Cambridge 
\bibitem{}van der Klis M., 1995, in {\it X-Ray Binaries}, eds. Lewin, van Paradijs \& van den Heuvel, Cambridge University Press: Cambridge
\bibitem{}van der Marel R.P., Anderson J., 2010, ApJ, 710, 1063
\bibitem{}Verbunt F., Hut P., 1987, IAUS, 125, 187
\bibitem{}Vitello P., Shlosman I., 1988, ApJ, 327, 680
\bibitem{}Watkins L.L., van de Ven G., den Brok M., van den Bosch R.C.E., 2013, MNRAS, 436, 2598
\bibitem{}Woosley S.E., Taam R.E., 1976, Nature, 263, 101
\bibitem{}Wheatley P.J., Mauche C.W., Mattei J.A., 2003, MNRAS, 345, 49
\bibitem{}White N.E., Marshall F.E., 1984, ApJ, 281, 354
\bibitem{}White N.E., Stellar L., Parmar A.n., 1988, ApJ, 324, 363
\bibitem{}Wu K., Soria R., Hunstead R.W., Johnston H., 2001, MNRAS, 320, 177
\bibitem{}Zhang S.N., Cui W., Chen W., 1997, ApJ, 482L, 155
\bibitem{}Zurek D.R., Knigge C., Maccarone T.J., Dieball A., Long K.S., 2009, ApJ, 699, 1113
\end{thebibliography}
\end{document}